# Microstructure.jl: a Julia Package for Probabilistic Microstructure Model Fitting with Diffusion MRI

Ting Gong[1*], Anastasia Yendiki[1]
1 Martinos Center for Biomedical Imaging, Massachusetts General Hospital and Harvard Medical School, Charlestown, MA, United States
*Correspondence: tgong1@mgh.harvard.edu

**Abstract:** Microstructure.jl is a Julia package designed for probabilistic estimation of tissue microstructural parameters from diffusion or combined diffusion-relaxometry MRI data. It provides a flexible and extensible framework for defining compartment models and includes robust and unified estimators for parameter fitting and uncertainty quantification. The package incorporates several established models from the literature, such as the spherical mean technique and soma and neurite density imaging (SANDI), along with their extensions for analyzing combined diffusion and T2 mapping data acquired at multiple echo times. For parameter estimation, it features methods like Markov Chain Monte Carlo (MCMC) sampling and Monte Carlo dropout with neural networks, which provide probabilistic estimates by approximating the posterior distributions of model parameters. In this study, we introduce the major modules, functionality, and design of this package. We demonstrate its usage in optimizing acquisition protocols and evaluating model fitting performance with synthesized datasets. We also showcase practical applications with publicly available datasets. Microstructure.jl is applicable to in vivo and ex vivo imaging data acquired with typical research, high-performance, or pre-clinical scanners.

# 1. Introduction

Microstructure modeling allows us to estimate biologically meaningful cellular parameters from MRI data, demonstrating high sensitivity and specificity in quantifying tissue morphology (Alexander et al., 2019; Novikov et al., 2019). This is achieved by parsimonious models that represent the tissue in a voxel as composed of a small set of compartments. Each tissue compartment corresponds to a different cellular structure, e.g., neurites or cell bodies, whose distinct geometry and composition gives rise to a different MR signal. Given data collected with an appropriate acquisition protocol, one can therefore disentangle the signals from these compartments and estimate their free parameters by solving an inverse problem. These parameters are informative on various tissue properties and can therefore serve as biomarkers for both healthy and diseased biological processes.

**1.1 Challenges & Developments**

There are several challenges involved in fitting complex microstructural models to MRI measurements that are inherently noisy and exhibit varying sensitivities to different tissue features. **(1) Presence of local minima**: Non-linear objective functions defined on high-dimensional parameter spaces often have multiple local minima. As a result, optimization algorithms may converge on suboptimal solutions that are inaccurate estimates of the underlying tissue properties, and that are highly dependent on how the algorithm was initialized. **(2) Parameter degeneracy**: Different combinations of tissue parameters can produce similar MRI measurements, making it difficult to distinguish between them based on the data alone, complicating the interpretation of parameter estimates. **(3) Lack of robustness**: Various factors, related to either experimental design or true individual variability, can degrade the accuracy of parameter estimates. On the one hand, MRI acquisition parameters, such as gradient strength or echo time (TE), affect both the sensitivity of the measurements to specific tissue features, and image quality characteristics such as signal-to-noise ratio (SNR) and artifacts. On the other hand, the heterogeneity of tissue properties across different regions or subjects can introduce estimation errors, if it is not modeled appropriately.

Maximum likelihood optimization methods, which provide a point estimate of parameters, are insufficient for quantifying and addressing these issues. For robust interpretation of the imaging measures, it is critical to provide estimates not only of the parameters of interest, but also of their uncertainty. Bayesian optimization methods allow uncertainty to be quantified by estimating the posterior distribution of parameters. This is also useful for identifying parameter degeneracy. For example, random sampling methods such as Markov Chain Monte Carlo (MCMC) generate samples drawn from the posterior distribution of the parameters (Gilks et al., 1995). These distributions capture the uncertainty around the estimate (Alexander, 2008; Behrens et al., 2003). The drawback of Bayesian methods is that they are computationally intensive, making them impractical for high-resolution, whole-brain datasets, and that they require tuning of the sampling parameters for each model.

Improving the speed and robustness of probabilistic microstructure model fitting is essential for making these models practical in large-scale neuroimaging studies. This involves a two-pronged approach: (i) leveraging high-performance languages and efficient computing to accelerate MCMC estimation methods and (ii) taking advantage of the recent developments in neural network-based estimation methods for even faster inference. Both MCMC and neural network-based approaches allow quantification of uncertainty in parameter estimates, and they may play complementary roles in managing the bias and variance trade-off of parameter estimation. Therefore, both methods are useful for addressing the challenges of microstructure imaging.

**1.2 Microstructure models**

Here, we introduce microstructure models briefly, to provide a common language for describing the features available in existing toolboxes and in Microstructure.jl.

**Orientation-dependent models**: This approach models the effects of the fiber orientations on the diffusion-weighted imaging (DWI) signal. The input data used for model fitting are either the DWI measurements collected with different gradient directions or their rotationally invariant features (Alexander et al., 2010; Assaf and Basser, 2005; Behrens et al., 2003; Jespersen et al., 2010, 2007; Novikov et al., 2018; Zhang et al., 2012).

**Orientation-averaged models**: This approach factors out the effects of fiber orientation through powder averaging (PA) (Callaghan et al., 1979) to focus on estimating other microstructure parameters. Thus, the input data used for model fitting are only PA signals obtained by averaging or fitting DWI measurements from different gradient directions at each unique b-value and/or diffusion time. Examples include the spherical mean technique (Kaden et al., 2016b), soma and neurite density imaging (SANDI) (Palombo et al., 2020), and several water exchange models (Jelescu et al., 2022; Olesen et al., 2022).

**Combined diffusion-relaxometry models**: These models extend those in the first two categories by introducing compartment-specific T2/T2* and/or T1 values as model parameters (Gong et al., 2023b, 2022, 2020; Lampinen et al., 2019; Slator et al., 2021; Veraart et al., 2017). They require DWI measurements to be collected at multiple echo times (TE) and/or inversion recovery times (TI). They may model fiber orientation explicitly or they may factor it out by PA. Whereas the two previous categories of models may yield relaxation-weighted and protocol-dependent estimates of tissue compartment fractions, combined diffusion-relaxometry models enable unbiased estimation. They also provide estimates of compartmental T2/T2*/T1 values that reflect the macromolecular environment of tissue compartments. While these models demand more extensive measurements and thus longer acquisition times, development of novel, efficient acquisition sequences (Dong et al., 2022; Fair et al., 2021; Hutter et al., 2018; Jun et al., 2024; Wang et al., 2019) and robust computation tools (de Almeida Martins et al., 2021; Gong et al., 2021a, 2021b; Palombo et al., 2023) is now making them practical for use in future studies.

**1.3 Summary of existing software**

**MATLAB toolboxes**. Several well-known toolboxes have made microstructure model fitting more accessible for basic and clinical neuroscience research. For example, the compartment modeling module in the UCL Camino Diffusion MRI Toolkit (Cook et al., 2005) implements a broad range of models, which have been described in a model comparison paper (Panagiotaki et al., 2012). These include models like ActiveAx for axon diameter estimation (Alexander et al., 2010) and VERDICT for cell size estimation (Panagiotaki et al., 2014). The NODDI toolbox, also from UCL, focuses on orientation-dispersed models for axon diameter (Zhang et al., 2011) and neurite density estimation (Tariq et al., 2016; Zhang et al.,

2012). These MATLAB toolboxes are user-friendly tools for applying several models from the literature in research settings.

**Python toolboxes**. There are several Python toolboxes designed to provide consistent frameworks for parameter estimation across various microstructure models. The Microstructure Diffusion Toolbox (MDT) (Harms and Roebroeck, 2018) implements several orientation-dependent models and provides MCMC sampling methods for estimating parameters and quantifying uncertainty. It also supports GPU processing to speed up model fitting. Dmipy (Fick et al., 2019) is another python toolbox that implements a wide range of tissue compartments, including both orientation-distributed and spherical mean-based models, and provides several non-linear solvers for parameter estimation. In addition to MDT and Dmipy, the Amico toolbox (Daducci et al., 2015) implements several models from the literature, including ActiveAx, NODDI, and SANDI, with accelerated processing using linear approximations.

**Other open-source tools.** While not included in previous sections, some studies have shared code for implementing their microstructure models. For example, the SANDI toolbox (Palombo et al., 2020), the standard model imaging toolbox (Coelho et al., 2022), and qMRINet (Grussu et al., 2021) provide tools to fit specific models with machine learning. Other software tools, such as MRtrix (Tournier et al., 2019) and Dipy (Garyfallidis et al., 2014), focus primarily on signal models for fiber orientation estimation and tractography.

**1.4 Features of Microstructure.jl**
Microstructure.jl aims to provide a flexible and probabilistic framework for microstructure model fitting, leveraging the intuitive and high-performance language Julia (Bezanson et al., 2014). It is designed not only as an application tool with a user-friendly interface but also as a developer tool that can be easily extended. Compared to existing toolboxes, it has a unique combination of features: 1. Support for combined diffusion-relaxometry modelling. 2. Generic MCMC and neural network estimators for parameter estimation and uncertainty quantification, in which model fitting assumptions can be easily adjusted. 3. Compatibility with other probabilistic programming packages in the Julia ecosystem, such as Turing.jl. Currently, Microstructure.jl includes spherical-mean based models for microstructure imaging. In parallel, we are developing a companion Julia package, Fibers.jl, for fiber orientation estimation and tractography.

In the following sections, we describe the main functionality and structure (section 2) and demonstrate key use cases (section 3). Microstructure.jl can be installed seamlessly through Julia's built-in package manager. Source code is available at: https://github.com/Tinggong/Microstructure.jl. Application Programming Interface (API) manuals and tutorials are available at: https://tinggong.github.io/Microstructure.jl/dev/.

## 2. Package design
**Figure 1** demonstrates an overview of the relationships between major data types and functions in Microstructure.jl. In this section, we introduce each module in detail. **Figure 2** includes code snippets that demonstrate key features and use cases introduced in sections 2 and 3.

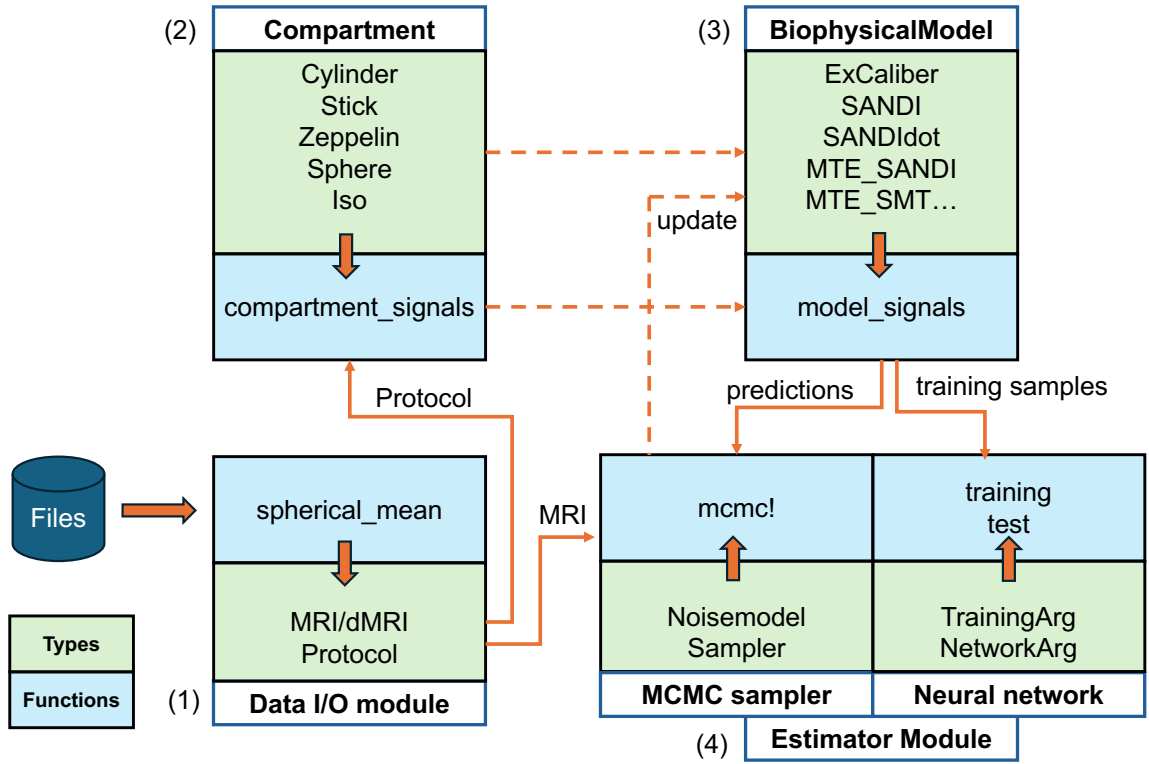

**Figure 1. Package overview.** The diagram shows the main data types (green boxes) and functions (blue boxes) in each module of Microstructure.jl. The direction of the orange arrow between a green and blue box indicates if a certain type is an input or output to the respective function.

### 2.1 Data I/O

In the I/O module, Microstructure.jl handles reading DWI volumes and text files with acquisition parameters to prepare the input data and protocols necessary for microstructure parameter fitting. The module includes functions to read DWI measurements from NIfTI or Bruker image files (via calls to the Fibers.jl package) and to perform gradient direction averaging for different b-values, TEs, diffusion gradient duration and separation times. These functions return a structure of type *'MRI'* that contains the volume series of spherical mean signals, and a structure of type *'Protocol'* that contains per-volume acquisition parameters such as diffusion times, and echo times that are necessary for advanced microstructure modeling. **Figure 2(i)** shows example code for reading data and generating spherical mean measurements and imaging protocol.

### 2.2 Tissue Compartments

In Microstructure.jl, tissue compartments are implemented as composite types, with fields that store relevant tissue microstructure parameters. The function *'compartment_signals'* predicts the normalized spherical mean signals (Callaghan et al., 1979; Kaden et al., 2016b; Kroenke et al., 2004) given a tissue compartment object and an imaging protocol. We take advantage of multiple dispatch in Julia to flexibly implement different methods (meaning the forward model) based on the compartment type. When specified, compartmental T2 values are included in the forward model, as demonstrated in previous work (Gong et al., 2023b, 2020; Lampinen et al., 2019; Veraart et al., 2017) for combined diffusion-relaxometry imaging. Currently, 5 different compartments are included in the package. Their forward models are given below. While we define all relevant parameters for each compartment, this does not mean that they can all be estimated robustly. The decision about which parameters to estimate should be made based on the acquisition protocol of the input data. Our framework allows the user to define explicitly which parameters will be estimated and which will be assumed to be fixed, and to pass any such assumptions to the estimator functions.

In all of the following model definitions, $b$ is the diffusion weighting b-value that can be computed from the diffusion gradient duration time $\delta$, separation time $\Delta$ and magnitude $G$, as $b = \gamma^2 G^2 \delta^2 (\Delta - \delta/3)$ for a pulsed gradient spin echo encoding; and $t$ is the TE. A relaxation weighting term $e^{-t/T_2}$ is included in each of these models for combined diffusion-relaxometry imaging only if multi-TE data (i.e., data acquired with different values of $t$) are provided and is otherwise omitted for conventional diffusion microstructure imaging. The forward model formular for each compartment below expresses the DWI signal at a voxel as a function of the unknown microstructural parameters in that voxel, as well as the known protocol variables $b$ and $t$. In a typical acquisition, multiple measurements of the signal in each voxel are acquired for different values of $b$ and possibly $t$, to allow estimation of the unknown microstructural parameters. In all expressions below, the DWI signal is assumed to be normalized by the b=0 signal in the same voxel.

**Zeppelin.** The zeppelin model is a tensor with equal water diffusivity in the two perpendicular directions, which is expected to be smaller than the water diffusivity in the parallel direction. This compartment is usually used to represent an anisotropic extra-cellular environment typically seen in the white matter (WM) (Alexander, 2008; Jespersen et al., 2010). The normalized signal follows the spherical mean expression as:

$$S_{zeppelin}(b(\delta, \Delta, G), \ t; \ D_\parallel, D_\perp, T_2) = e^{-t/T_2} e^{-bD_\perp} \sqrt{\frac{\pi}{4b(D_\parallel - D_\perp)}} \ \text{erf}\left(\sqrt{b(D_\parallel - D_\perp)}\right) \ [1],$$

where the three free parameters $D_\parallel$, $D_\perp$, and $T_2$ are the parallel diffusivity, perpendicular diffusivity, and the T2 value of water within the Zeppelin compartment; and erf($\cdot$) is the error function. To constrain $D_\perp$ to be smaller than $D_\parallel$, $D_\perp$ is represented as $D_\parallel$ multiplied by a fraction between [0, 1].

**Cylinder.** In the cylinder model, water diffusion in the perpendicular direction in Eq [1] is related to the cylinder diameter through the Gaussian phase distribution (GPD) approximation. (Andersson et al., 2022; Fan et al., 2020; Van Gelderen et al., 1994). This compartment is used to represent axons when the imaging acquisition protocol can provide sensitivity to the axon diameter. Therefore, the normalized signal is:

$$S_{cylinder}(b(\delta, \Delta, G), \ t; \ D_\parallel, D_0, d_a, T_2) = e^{-t/T_2} e^{-bD_\perp} \sqrt{\frac{\pi}{4b(D_\parallel - D_\perp)}} \ \text{erf}\left(\sqrt{b(D_\parallel - D_\perp)}\right);$$

And $-bD_\perp =$
$$-2\gamma^2 G^2 \sum_{m=1}^{\infty} \frac{1}{D_0^2 \alpha_m^6 (r_a^2 \alpha_m^2 - 1)} \left(2 D_0 \alpha_m^2 \delta - 2 + 2e^{-D_0 \alpha_m^2 \delta} + 2e^{-D_0 \alpha_m^2 \Delta} - e^{-D_0 \alpha_m^2 (\Delta - \delta)} - e^{-D_0 \alpha_m^2 (\Delta + \delta)}\right) \ [2],$$

where the four free parameters $D_\parallel, D_0, d_a$, and $T_2$ are the parallel diffusivity, intrinsic diffusivity, cylinder diameter, and T2 value, respectively; $r_a = d_a/2$ is the cylinder radius; $\alpha_m$ is the $m$-th root of $J_1'(\alpha r_a) = 0$ and $J_1'$ is the derivative of the first-order Bessel function of the first kind. Because cylinder diameter is known to be a biased representation of axon diameter (Burcaw et al., 2015; Veraart et al., 2020), we refer to $d_a$ as *axon diameter index* in the rest of the manuscript.

**Stick.** When the imaging protocol is not sensitive to the axon diameter index, a stick model is used to represent the axons (Behrens et al., 2003), where the water diffusivity perpendicular to the axons is approximately zero, and therefore the normalized signal is given by:

$$S_{stick}(b(\delta, \Delta, G), t;\ D_\|, T_2) = e^{-t/T_2} \sqrt{\frac{\pi}{4bD_\|}}\ \text{erf}(\sqrt{bD_\|})\ [3],$$

where the two free parameters $D_\|$ and $T_2$ and the parallel diffusivity and T2 values of water, respectively, in the stick compartment. This compartment is used for estimating the intra-neurite signal fraction and diffusivity with lower b-value acquisition protocols.

**Sphere.** The sphere model is an isotropic compartment where diffusion of water molecules is assumed to be restricted in spheres of radius $r_s$. This compartment is used to represent cell bodies more abundant in the gray matter (GM) (Palombo et al., 2020). The normalized signal given by GPD is (Balinov et al., 1993; Neuman, 1974):

$$S_{sphere}(b(\delta, \Delta, G), t;\ D_{is}, r_s, T_2) =$$
$$e^{-t/T_2} e^{\{-2\gamma^2 G^2 \sum_{m=1}^{\infty} \frac{1}{D_{is}^2 \alpha_m^6 (r_a^2 \alpha_m^2 - 2)}(2D_{is}\alpha_m^2 \delta - 2 + 2e^{-D_{is}\alpha_m^2 \delta} + 2e^{-D_{is}\alpha_m^2 \Delta} - e^{-D_{is}\alpha_m^2 (\Delta - \delta)} - e^{-D_{is}\alpha_m^2 (\Delta + \delta)})\}}\ [4],$$

where the three free parameters $D_{is}, r_s$, and $T_2$ are the water diffusivity, radius, and T2 value, respectively, in the sphere compartment; $\alpha_m$ is the $m$-th root of $\frac{1}{\alpha r_s} J_{\frac{3}{2}}(\alpha r_s) = J_{\frac{5}{2}}(\alpha r_s)$ and $J_n$ is the Bessel function of the first kind.

**Iso.** This compartment is used to model tissue environments in which water exhibits isotropic Gaussian diffusion. Its normalized signal is thus given by:
$$S_{iso}(b(\delta, \Delta, G), t;\ D, T_2) = e^{-t/T_2} e^{-bD}\ [5],$$
where the free parameters $D$ and $T_2$ are diffusivity and T2 value of water, respectively, in the compartment. An isotropic compartment can be used to model: (i) cerebrospinal fluid (CSF), where $D$ is free water diffusivity; (ii) the extra-cellular compartment, where $D$ is dependent on the tissue environment and estimated from the data; or (iii) immobile water in ex vivo tissue, where the diffusivity $D$ is set to 0.

### 2.3 Biophysical Models

A biophysical model for microstructure imaging is a representation of the signals measured in a tissue voxel by a linear combination of contributions from multiple compartments. In Microstructure.jl, water exchange between compartments is currently not considered; therefore, one should consider whether the experimental regime is appropriate for this assumption. This section introduces examples specific to modeling GM and WM tissue.

**GM models.** The package includes the SANDI model (Palombo et al., 2020) and its extensions for GM microstructure modeling. The *'SANDI'* model is a three-compartment model, where cell soma, neurite, and extra-cellular space are modeled as *'Sphere'*, *'Stick'*, and *'Iso'* compartments, respectively. For ex vivo imaging, the *'SANDIdot'* model includes an additional "dot" compartment (*'Iso'* compartment with diffusivity equal to 0) to account for immobile water in ex vivo tissue (Olesen et al., 2022). We have used this model to study cellular changes due to brain development in ex vivo macaque brain samples from early infancy to adulthood (Gong et al., 2024). Lastly, the *'MTE_SANDI'* model supports data collected with multiple echo times and can also estimate the T2 values in the soma, neurite, and extra-cellular compartments (Gong et al., 2023b).

**WM models.** The package includes two types of WM models. The first type models the axonal space as a *'Cylinder'* compartment and assumes that the data are acquired with strong enough diffusion gradients to be sensitive to the size of axons (Andersson et al., 2022; Fan et al., 2020). The *'ExCaliber'* model is one example, where intra-axonal space is modeled as a *'Cylinder'* compartment, extra-cellular space modeled as a *'Zeppelin'* compartment, and an *'Iso'* compartment is used either with diffusivity equal to zero, to represent immobile water when imaging ex vivo tissue (Gong et al., 2025), or with diffusivity equal to that of free water, to represent CSF when imaging in vivo tissue (Fan et al., 2020). The second type of WM model represents axons as a *'Stick'* compartment and is applicable to data acquired on clinical MRI systems that do not allow high enough b-values to sensitize the signal to axon diameters (Kaden et al., 2016b). *'MTE_SMT'* is such a model, with *'Stick'* and *'Zeppelin'* compartments that have distinct compartmental T2 values.

The function *'model_signals'* predicts the normalized spherical mean signals given a model object and an imaging protocol, similar to the *'compartment_signals'* function. **Figure 2(ii)** shows example code for generating signals using a WM tissue model.

## 2.4 Estimators

Estimators in Microstructure.jl provide flexible settings to determine which parameters of a biophysical model will be estimated and how. Given that the available data may not be suitable for robust estimation of all free parameters in a model, this flexibility allows the evaluation and adjustment of model fitting assumptions. Details about the estimators can be found in our API manuals on the documentation website. Below we provide brief overviews and considerations for choosing between the two types of estimator.

**MCMC.** Microstructure.jl implements the Metropolis-Hastings sampling algorithm without relying on external sampling libraries. It provides flexible arguments through the *'Sampler'* and *'Noisemodel'* inputs, allowing the user to apply the estimator to sample any parameters from any biophysical model. It is also possible to apply other advanced MCMC sampling methods to models in Microstructure.jl with Turing.jl (https://turinglang.org/), a general-purpose library for Bayesian inference. Initial evaluations indicate that Turing.jl sampling is slower than the native MCMC implementation in Microstructure.jl, and more detailed testing will be conducted in the future. To accelerate MCMC sampling for large datasets, Microstructure.jl supports multi-thread processing.

**MC dropout with neural networks.** The neural network estimator in Microstructure.jl builds a multi-layer perceptron (MLP) model and trains the model with synthetic training data for a specific *'BiophysicalModel'*. The training data contain pairs of model parameters and predicted measurements given the measurement protocol, noise model, and the noise level for supervised learning. Dropout layers are used as a Bayesian approximation to estimate the parameter posterior distributions (Gal and Ghahramani, 2015). This module uses the machine learning library Flux.jl (https://fluxml.ai/) and allows flexible parameterization of MLPs through *'NetworkArg'* and training options through *'TrainingArg'*. While simple network architectures can be effective for microstructure fitting, the accuracy and generalizability of trained network models is highly dependent on the training datasets. For example, training samples generated by fitting microstructural models on real brain datasets can introduce unbalanced prior distributions of the parameters, depending on the population that the training data were acquired from. This can introduce bias to estimates, particularly in regions where the true parameter values in the test data are underrepresented in the training set (Li et al., 2019).

**Considerations for choosing an estimator. (1) Speed**. The neural network estimators are much faster. A whole-brain dataset can typically be processed in minutes on a single central processing unit (CPU) across different models, making neural network estimators convenient for processing large-scale datasets. The processing time for the MCMC sampler, on the other hand, increases linearly with the size of the data, thus is much slower and may become prohibitive when processing high-resolution, whole-brain datasets. For acceleration, the MCMC sampler currently supports multi-thread processing on CPUs. **(2) Priors.** The MCMC estimator uses uniform prior distributions within given parameter ranges. In contrast, the neural network estimator allows the incorporation of any prior distributions through synthetic training datasets. Uniform priors yield estimates that are less biased by prior knowledge, while informative priors could improve parameter estimation if such prior knowledge is valid and effective.

In brief, the classical MCMC estimator is helpful for evaluating protocol and fitting at the single voxel level, while the neural network estimator offers faster processing speed and greater flexibility for introducing prior information to improve parameter estimation.

**Snippet i: Reading data and protocol**
```
dmri, protocol, snr = spherical_mean(
    joinpath(datadir, "dwi.nii.gz"), # DWI data
    true,                            # save normalized data too?
    joinpath(datadir, "dwi.bvals"), joinpath(datadir, "dwi.bvecs"),
    joinpath(datadir, "dwi.techo"), joinpath(datadir, "dwi.tdelta"),
    joinpath(datadir, "dwi.tsmalldel"))
```

**Snippet ii: Generating signals from compartment model**
```
model = ExCaliber(axon=Cylinder(da=2.0e-6),
    extra=Zeppelin(dperp_frac = 0.3), fracs = [0.7, 0.15])
signals = model_signals(model, prot)
```

**Snippet iii: Setting up MCMC estimator**
```
# set the tissue parameters you want to estimate in the model;
paras = ("axon.da","axon.dpara","extra.dperp_frac","fracs")
# set parameter links
paralinks = ("axon.d0" => "axon.dpara", "extra.dpara" => "axon.dpara")

# set the range of priors and proposal distributions
pararange = ((1.0e-7,1.0e-5),(0.01e-9,1.2e-9),(0.0, 1.0),(0.0,1.0))
proposal = (Normal(0,0.25e-6),Normal(0,0.025e-9),Normal(0,0.05),
    MvNormal([0.0025 0;0 0.0001]))

# setup two-stage sampler
sampler_full = Sampler(params=paras,prior_range=pararange,
    proposal=proposal,paralinks=paralinks,nsamples=70000,burnin=20000)
sampler_sub = subsampler(sampler_full,[1,4])
sampler = (sampler_full,sampler_sub)
```

**Snippet iv: Using neural network estimator for SMT**
```
### Setup model and nn estimator
# t2 is set to default as 0, so single-TE model will be used
model = MTE_SMT(
    axon = Stick(dpara = 2.0e-9, t2 = 0.0),
    extra = Zeppelin(dpara = 2.0e-9, dperp_frac = 0.5, t2 = 0.0),
    fracs = 0.5,
)

# parameters to estimate
params = ("fracs", "axon.dpara", "extra.dperp_frac")
prior_range = ((0.0, 1.0), (1.0e-9, 3.0e-9), (0.0, 1.0))
prior_dist = (nothing, nothing, nothing)
paralinks = ("extra.dpara" => "axon.dpara")
noise_type = "Gaussian"
sigma_range = (0.002, 0.02)
sigma_dist = Normal(0.01, 0.002)

nsamples = 30000
nin = 4
nout = 3
hidden_layers = (32, 32, 32)
dropoutp = (0.1, 0.1, 0.1)

netarg = NetworkArg(model, protocol, params, prior_range, prior_dist,
    paralinks, noise_type, sigma_range, sigma_dist, nsamples,
    nin, nout, hidden_layers, dropoutp, relu6)
trainarg = TrainingArg(batchsize = 128, lossf=losses_rmse, device = cpu)

# get trained model, training log and training data
mlp, logs, inputs, labels = training(trainarg, netarg)
```

**Snippet v: Using informative priors for SMT**
```
# parameters to estimate
params = ("fracs", "axon.dpara", "extra.dperp_frac")
prior_range = ((0.0, 1.0), (1.0e-9, 3.0e-9), (0.0, 1.0))
prior_dist = (nothing, Normal(2.0e-9, 0.3e-9), nothing)
paralinks = ("extra.dpara" => "axon.dpara")
```

**Snippet vi: Using neural network estimator for SANDI**
```
### 2. Setup model and estimator
model = SANDI(soma = Sphere(diff = 3.0e-9, size = 8.0e-9),
    neurite = Stick(dpara=2.0e-9),
    extra = Iso(diff = 2.0e-9),
    fracs = [0.4,0.4] )

# parameters to estimate
params = ("fracs", "soma.size", "neurite.dpara", "extra.diff")
prior_range = ((0.0, 1.0), (2.0e-6, 12.0e-6), (1.5e-9, 2.5e-9),
    (0.5e-9, 3.0e-9))
prior_dist = (Dirichlet([1,1,1]), nothing, nothing, nothing)
paralinks = ()
noise_type = "Gaussian"
sigma_range = (0.002, 0.02)
sigma_dist = Normal(0.006, 0.002)

# network settings
nsamples = 60000
nin = 8
nout = 5
hidden_layers = (48, 48, 48)
dropoutp = (0.1, 0.1, 0.1)
```

**Figure 2. Code snippets that demonstrate example usage of the package. (i)** Reading DWIs and acquisition parameters and returning the direction-averaged measurements and the associated protocol structure. (The acquisition parameter files that include b-values and optional gradient directions, echo times, diffusion gradient pulse separation and width are assumed to have the specified extensions.) SNR maps are also returned as the ratio of mean and standard deviation of *b*=0 measurements at each echo

time. **(ii)** Generating signals from a compartment or tissue model. **(iii)** Setting up an MCMC sampler using the two-stage method as an example. **(iv)** Setting up a neural network estimator and training a network for SMT estimation. **(v)** Changing the prior distributions used for generating training datasets in (iv). **(vi)** Setting up a neural network estimator for SANDI estimation. Functionality may evolve as the package is developed – please refer to the package documentation website for the latest information and more details.

3. **Demonstrations**

We demonstrate applications of Microstructure.jl focusing on (1) Optimizing acquisition protocols and evaluating quality-of-fit: We show how to generate signals from a single tissue compartment (section 3.1) and how to evaluate fitting strategies for multi-compartment models on simulated data (section 3.2). (2) Applying models from the literature on publicly available datasets: We show how to use neural network estimators for WM and GM models and adapt the prior distributions in training datasets (section 3.3). These demonstrations highlight the importance of having an easy-to-use package for optimizing data acquisition and evaluating parameter fitting for microstructure imaging. Lastly, we summarize the computation times for these tasks to demonstrate the performance of the package (section 3.4).

**3.1 Generating signals from compartment models**

Generating signals from a compartment model with a given imaging protocol and associated tissue properties is an essential building block of microstructure modeling. Before embarking on an imaging study, this functionality is key for understanding how the data should be collected to improve sensitivity to the tissue properties of interest. We demonstrate this for the task of designing an imaging protocol for estimating axon diameter index with the '*Cylinder*' compartment.

**3.1.1 Sensitivity range of axon diameter index**

When estimating axon diameter indices from MRI data, the range of values that we are sensitive to depends both on experimental factors such as gradient strength and diffusion times, and on tissue properties such as the intrinsic diffusivity. Therefore, evaluation of experimental settings is important for understanding the validity of the axon diameter index. Following the sensitivity criterion for the PA signal introduced by previous studies (Andersson et al., 2022; Nilsson et al., 2017), we can conveniently calculate the sensitivity range of axon diameter index estimates obtained from Microstructure.jl.

The sensitivity criterion concerns whether the normalized signal along a single gradient can be differentiated from the noise floor and the maximum signal. The smallest measurable signal attenuation $\bar{\sigma}$ relates to the SNR level of the signal as follows:

$$\bar{\sigma} = \frac{z_\alpha}{SNR\sqrt{n}},$$

where n is the number of measurements and $z_\alpha$ is the z-threshold for the significance level $\alpha$ ($z_\alpha$ = 1.64 when $\alpha = 0.05$). The upper bound of the PA signal from a '*Cylinder*' compartment is that of a '*Cylinder*' with zero diameter, which is equal to the '*Stick*' model signal. A '*Cylinder*' therefore has a measurable diameter if its PA signal falls within the range of [$\bar{\sigma}$, $S_{stick} - \bar{\sigma}$]. Given a single b-value, the lower and upper bounds of axon diameter are thus, respectively, the minimal diameter that gives $S_{cylinder}$ smaller than $S_{stick} - \bar{\sigma}$ and the maximal diameter that gives $S_{cylinder}$ above $\bar{\sigma}$.

**Experiments.** Using this criterion, we calculate the sensitivity ranges of axon diameter estimation at different high b-values feasible on a 4.7 T preclinical scanner with maximum gradient strength $G_{max}$ = 660 mT/m. We investigate the effects of diffusion time and b-value to sensitivity ranges by using different diffusion times that reach the $G_{max}$ with different b-values. We assume that the number of gradient directions is 32 and the ex vivo intrinsic diffusivity is 0.6 $\mu m^2$/ms. We consider SNR levels of 100, 50 and 30 in the *b*=0 measurement.

**Results.** For the same diffusion time, the sensitivity range shifts towards smaller axons when the gradient strength *G* and thus b-value increases, as seen in **Figure 3(A-C)**. Higher gradient strengths yield sensitivity to smaller axons, which are most abundant in tissue. However, they also reduce the sensitivity to very large axons, e.g. axons with diameter > 7.5 $\mu m$ for *G* = 660 mT/m. For the same gradient strength *G*, shorter diffusion time and therefore lower b-value widens the sensitivity range on both sides. For example, see *G*≈ 660 mT/m, *b* = 25, 43 and 64 in **Figure 3(A-C)**. For the same b-value, shorter diffusion time and thus higher *G* lowers the lower bound of axon diameters that can be estimated accurately but also reduces sensitivity to larger axons. This analysis suggests that using higher gradient strength and shorter diffusion time to achieve lower b-values is preferable, as it achieves a wider sensitivity range that covers most axon diameters expected in real tissue.

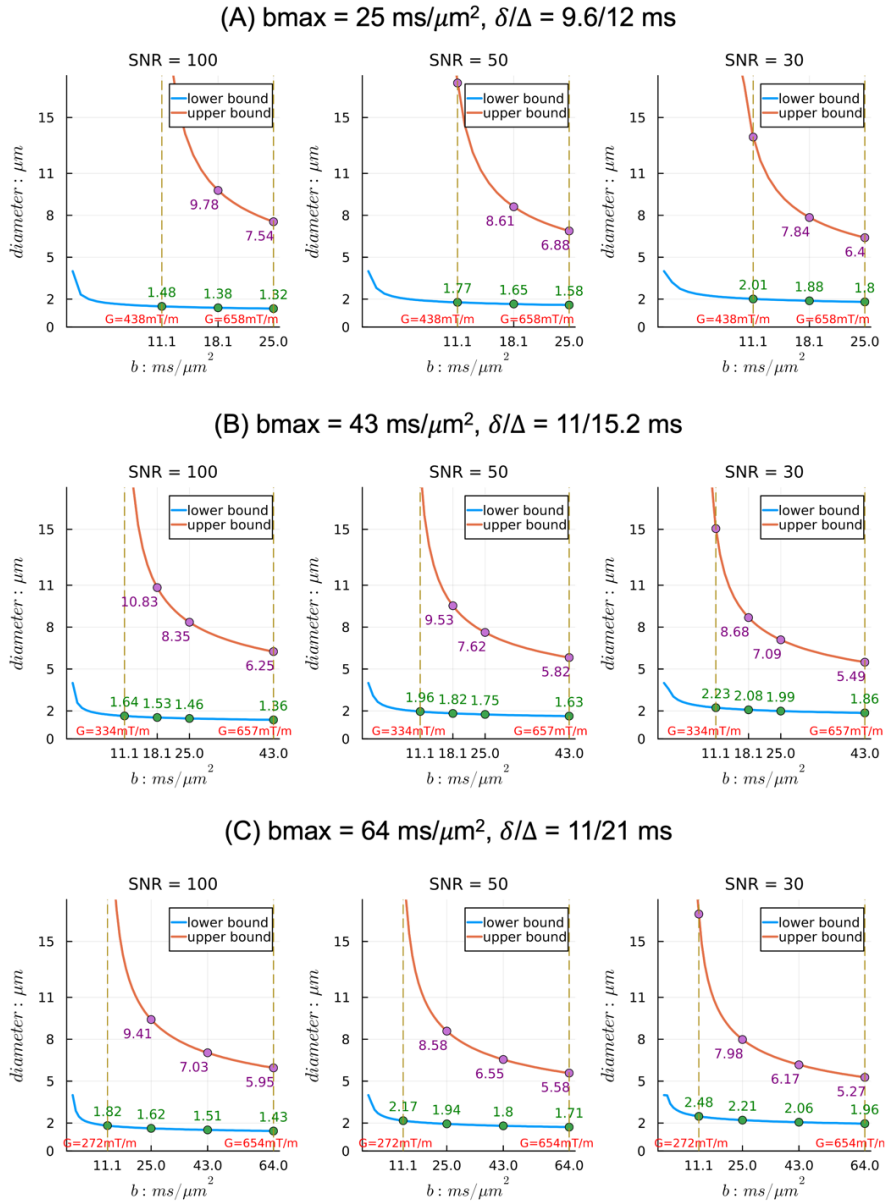

**Figure 3. Sensitivity ranges for axon diameter index estimation** from single b-value data with the 'Cylinder' model, assuming $D_0$ = 0.6 $\mu m^2$/ms, n = 32 directions, and different SNR levels in the b=0 measurements. Different diffusion times (A-C) are considered to achieve different b-values with Gmax ≈ 660 mT/m.

**Comparison to in vivo sensitivity ranges.** We can also assess the effects of the intrinsic diffusivity of tissue on the sensitivity range. We consider diffusivities typical of ex vivo brain ($D_0$ = 0.6 $\mu m^2$/ms) and in vivo brain ($D_0$ = 1.7 $\mu m^2$/ms). We include acquisition parameters relevant for in vivo human imaging with strong gradients (b = 9 ms/$\mu m^2$, G ~ 300 mT/m), in addition to the acquisition parameters relevant to ex vivo imaging that were detailed above. Results are shown in **Figure 4(B)**. The lower diffusivity plays a critical role in the lower resolution limits in ex vivo tissue.

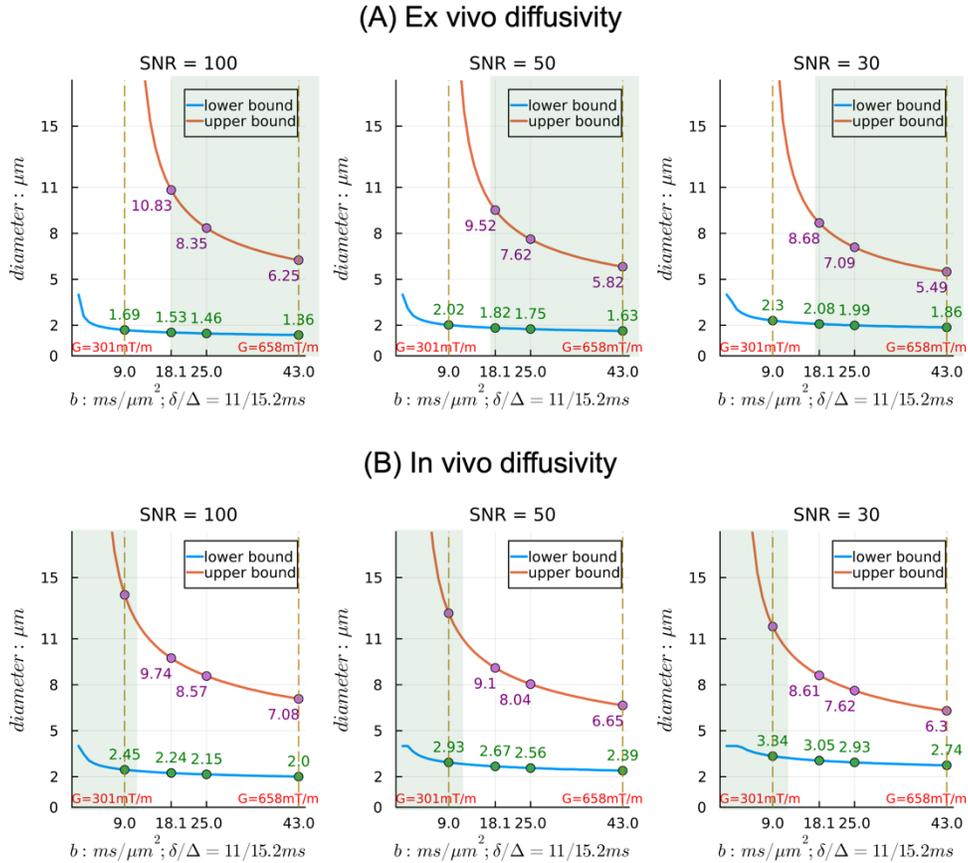

**Figure 4. Sensitivity ranges for axon diameter estimation** in ex vivo (A, $D_0$ = 0.6 $\mu m^2$/ms) and in vivo (B, $D_0$ = 1.7 $\mu m^2$/ms) brain tissue from single b-value data using the 'Cylinder' model, assuming n = 32 directions and different SNR levels in the b=0 measurements.

**Sensitivity range for multiple high b-values.** The above analysis assumes measurements with a single b-value. If data are collected with multiple b-values that are sensitive to the practical range of axon diameter in tissue, the lower and upper bounds of $S_{cylinder}$ (corresponding to the upper and lower bounds of axon diameter) become vectors $[\overline{\sigma}, S_{stick} - \overline{\sigma}]$. In this case, we can approximate the sensitivity range by finding the diameters that give minimal mean squared error (MSE) between their signals and signal bounds across all shells. For estimating the upper bound of axon diameters, we use only shells of high b-values because data with lower b-values can be sensitive to infinitely large axons, and if included their signal contributions will dominate the calculation of MSE. Considering the three high b-shells in **Figure 3(B)** for ex vivo imaging (b= 18.1, 25 and 43 ms/ $\mu m^2$), the sensitivity ranges are [1.42, 10.5], [1.70, 9.18], and [1.94, 8.32] $\mu m$ for SNR level of 100, 50 and 30 respectively.

### 3.2 Evaluating the quality-of-fit for a given model fitting strategy
After fitting a microstructure model, it is important to assess the quality of the fitting by inspecting the parameter posterior samples and determine if the fitting assumptions should be adjusted to improve fitting quality. We demonstrate this process with a two-stage MCMC fitting method that we developed for estimating axon diameter indices in ex vivo tissue (Gong et al., 2025). **Figure 2(iii)** shows example code for setting up such a two-stage MCMC sampler in Microstructure.jl.

### 3.2.1 Inspecting quality of fitting and posterior samples

For estimating axon diameter index in *ex vivo* tissue, previous studies have used only the intra-axonal compartment with very few, ultra-high b-values (>=20 ms/$\mu m^2$ for ex vivo tissue). We use the multi-compartment model *'ExCaliber'*, with the additional consideration that a nonnegligible and spatially varying dot signal is present in ex vivo tissue at high b-values and needs to be differentiated from the intra-axonal signals. We therefore model the full signal decay from a range of low and high b-values. The five free parameters in the *'ExCaliber'* model are the axon diameter index, the intra-axonal parallel diffusivity, the extra-cellular perpendicular diffusivity, and the intra-axonal and dot signal fraction. Parallel diffusivities in the intra-axonal and extra-cellular space are assumed to be equal to the intrinsic diffusivity.

We demonstrate this by simulating an example set of spherical mean measurements. This is done by computing the forward model for ground-truth parameters typical of a WM voxel: axon diameter index $d_a$ = 2 $\mu$m, intra-axonal signal fraction $f_{ia}$ = 0.7, dot signal fraction $f_{dot}$ = 0.15, parallel diffusivity $D_\parallel$ = 0.6 $\mu m^2$/ms and perpendicular diffusivity $D_\perp$ = 0.6 * 0.3 $\mu m^2$/ms. We add Gaussian-distributed noise to the spherical mean signals to generate measurements at SNR =100. **Figure 5** shows the quality of fit and posterior samples after the first and second MCMC run on these measurements. In the first stage, where all five tissue parameters are sampled, we find high uncertainty of estimated intra-axonal fractions. By fixing the parallel diffusivity and extra-cellular perpendicular diffusivity to their posterior means and sampling only the distributions of other 3 tissue parameters, the second MCMC run achieves similar likelihood of measurements, but lower parameter uncertainty and higher accuracy for the axon diameter index and compartment signal fractions.

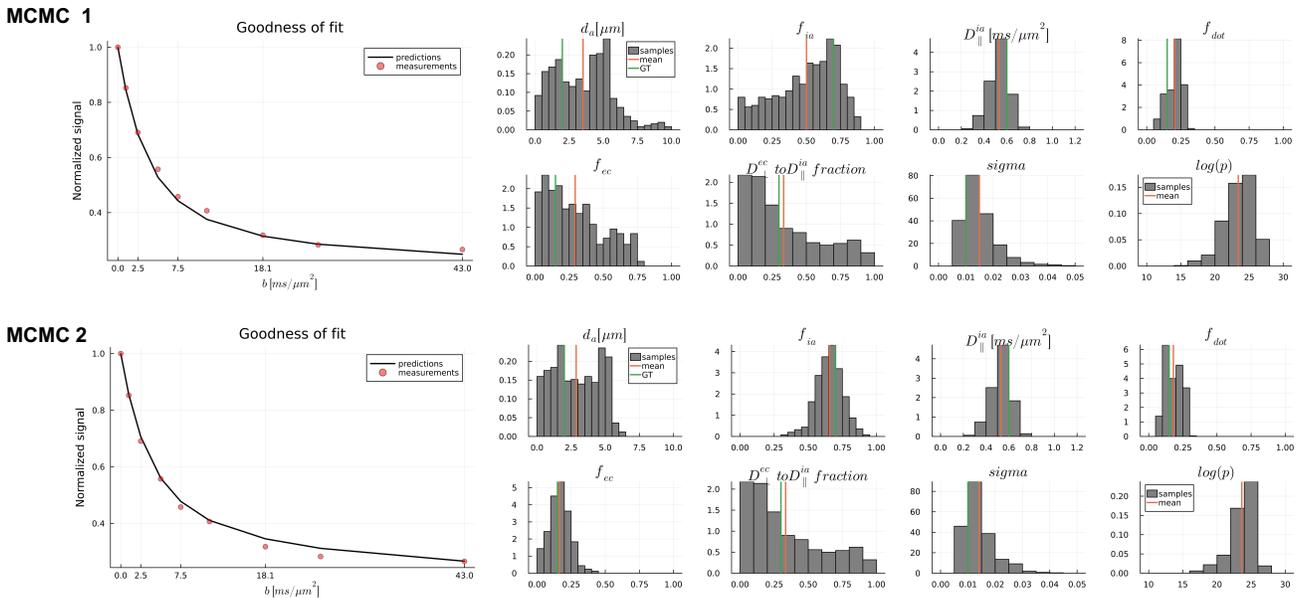

**Figure 5. Fitting curves and posterior samples from the two MCMC runs of a two-stage MCMC fitting approach.** We show the posterior distributions as histograms, with the mean estimates and ground truth (GT) tissue parameters indicated by green and orange lines, respectively. The parameters are the axon diameter index ($d_a$/ µm), the intra-axonal signal fraction ($f_{ia}$), the intra-axonal parallel diffusivity ($D_\parallel$/ µm$^2$/ms), the dot signal fraction ($f_{dot}$), the extra-cellular signal fraction ($f_{ec}$), the extra-cellular perpendicular diffusivity represented as a fraction to the parallel diffusivity, the noise level (sigma), and the log likelihood of measurements (log(p)). The histograms are normalized so that the area under each histogram is one.

### 3.2.2 Evaluation of estimation accuracy and precision

Synthetic datasets with known ground-truth parameters are essential for evaluating the accuracy and precision of microstructure model estimates for different data acquisition protocols. We demonstrate this approach for evaluating axon diameter index estimation, by generating signals using the three-compartment tissue model *'ExCaliber'* with axon diameter indices ranging from 1 µm to 10 µm at 1 µm intervals. This range of evaluation extends slightly beyond the sensitivity limits calculated from the *'Cylinder'* compartment in section 3.1.

The tissue parameters chosen for the synthetic data were values typically observed in ex vivo WM tissue: intra-axonal signal fraction $f_{ia}$ = 0.7, dot signal fraction $f_{dot}$ = 0.15, extra-cellular signal fraction $f_{ec}$ = 0.15, parallel diffusivity $D_\parallel$ = 0.6 µm$^2$/ms and perpendicular diffusivity $D_\perp$ = 0.6 * 0.3 µm$^2$/ms. We tested three protocols with a single diffusion time and multiple b-values, where the maximum b-value was chosen to reach $G_{max}$ = 660 mT/m. The first had δ/Δ = 9.6/12 ms and *b* = 1, 2.5, 5, 7.5, 11.1, 18.1, 25 ms/ µm$^2$ (7 b-values total), the second had δ/Δ = 11/15.192 ms and an additional *b* = 43 ms/ µm$^2$ (8 b-values total), and the third had δ/Δ = 11/21 ms and an additional *b* = 64 ms/ µm$^2$ (9 b-values total). Gaussian-distributed noise was added to generate 100 realizations of noisy spherical mean measurements; we show results for a moderate and lower SNR of 100 and 50 for the spherical mean measurements. **Figure 6-7** shows the estimated parameters using the two-stage approach of section 3.2 from the 100 noise realizations as boxplots and the ground truth values as line plots.

**Measurements with a single diffusion time (Figure 6).** Smaller axons had more precise diameter estimates and more accurate intra-axonal signal fractions than larger axons. However, diameter estimates in the low range were biased in a way that made different diameters more difficult to discriminate. In comparison, the diameters and intra-axonal signal fractions of larger axons were always under-estimated, while the dot signal fractions were more accurate. Among the different diffusion times, the shorter ones (Δ = 12 and 15.2 ms) maintained a consistent trend of axon diameter index estimates within the resolution limit (about 2-8 $\mu$m). Comparison of the two SNR levels shows that the lower SNR (SNR = 50) decreased estimation precision for smaller axons but increased the discriminability of axon diameter indices in the low range. The lower SNR also increased bias for larger axons. These findings highlight the importance of performing such evaluations to understand the effects of different acquisition parameters on model fitting.

**Measurements with multiple diffusion times (Figure 7).** Including data with multiple diffusion times improved the accuracy of axon diameter estimation, with better discriminability between smaller axons and lower biases and variances for larger axons, particularly at high SNR. However, at lower

SNR, there was less to be gained by including multiple diffusion times vs. a single diffusion time (see **Figure 6(B,i-ii)** and **Figure 7(B,i-ii))**, as the differences between signals with different diffusion times were very small. In real datasets, we need to consider if the SNR is sufficient for the differences between signals acquired with different diffusion times to be significant. When using data from all the diffusion times for fitting, both the bias and variance decreased substantially.

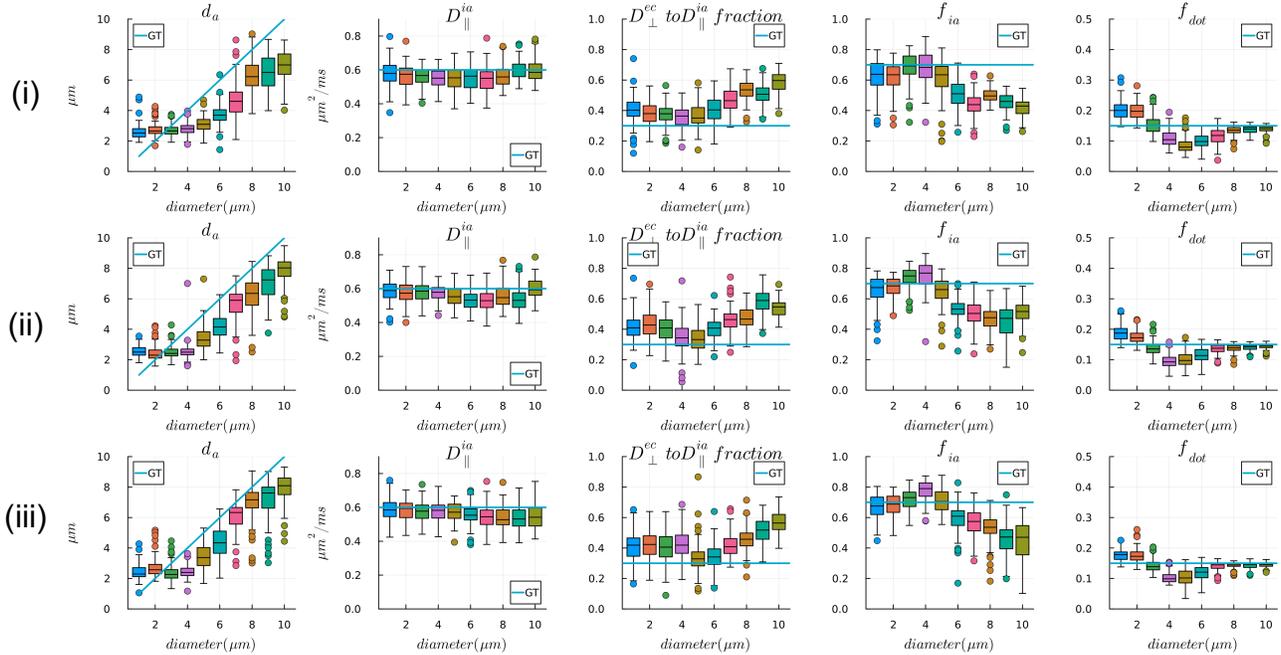

(A) Data of single diffusion time, SNR = 100

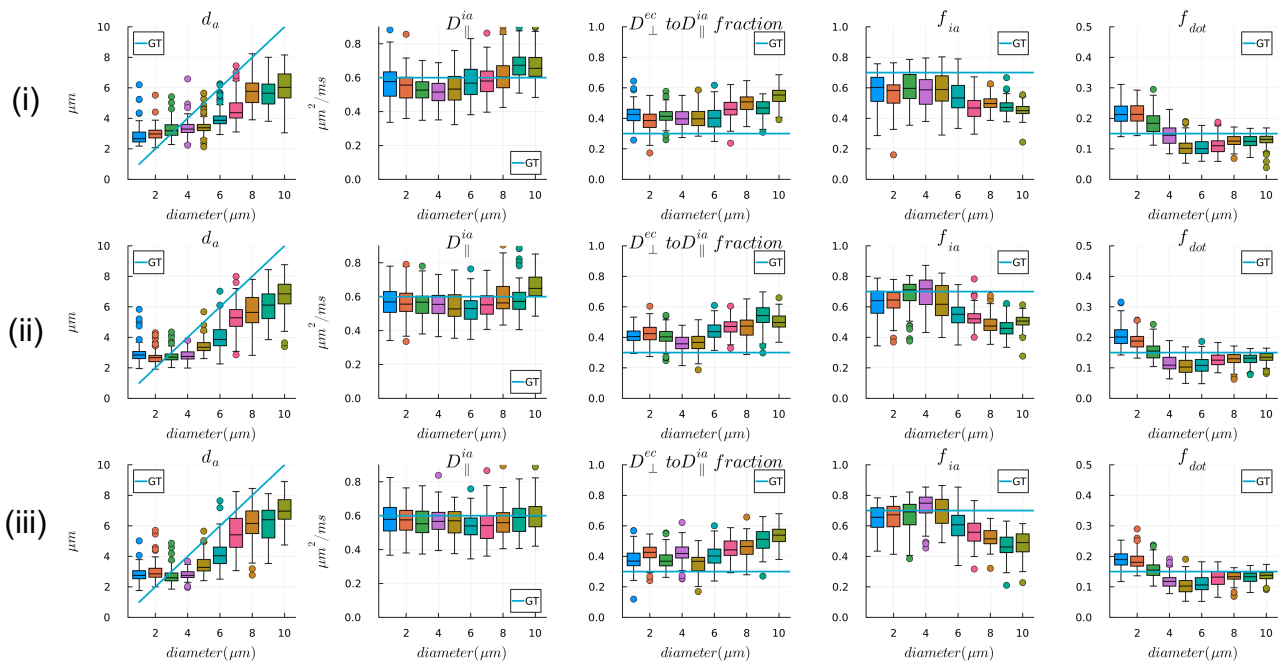

(B) Data of single diffusion time, SNR = 50

**Figure 6. Estimates of axon diameters from single diffusion time data** when (A) SNR = 100 and (B) SNR = 50 for spherical mean signal. (i) Data generated with 7 values, $b_{max}$ = 25 ms/µm$^2$ and $\delta/\Delta$ = 9.6/12 ms; (ii) Data generated with 8 b-values, $b_{max}$ = 43 ms/µm$^2$ and $\delta/\Delta$ = 11/15.2 ms; (iii) Data generated with 9 b-values, $b_{max}$ = 64 ms/µm$^2$ and $\delta/\Delta$ = 11/21 ms. Parameter estimates from 100 noise realizations are shown as boxplots and ground-truth (GT) parameter values are shown as line plots.

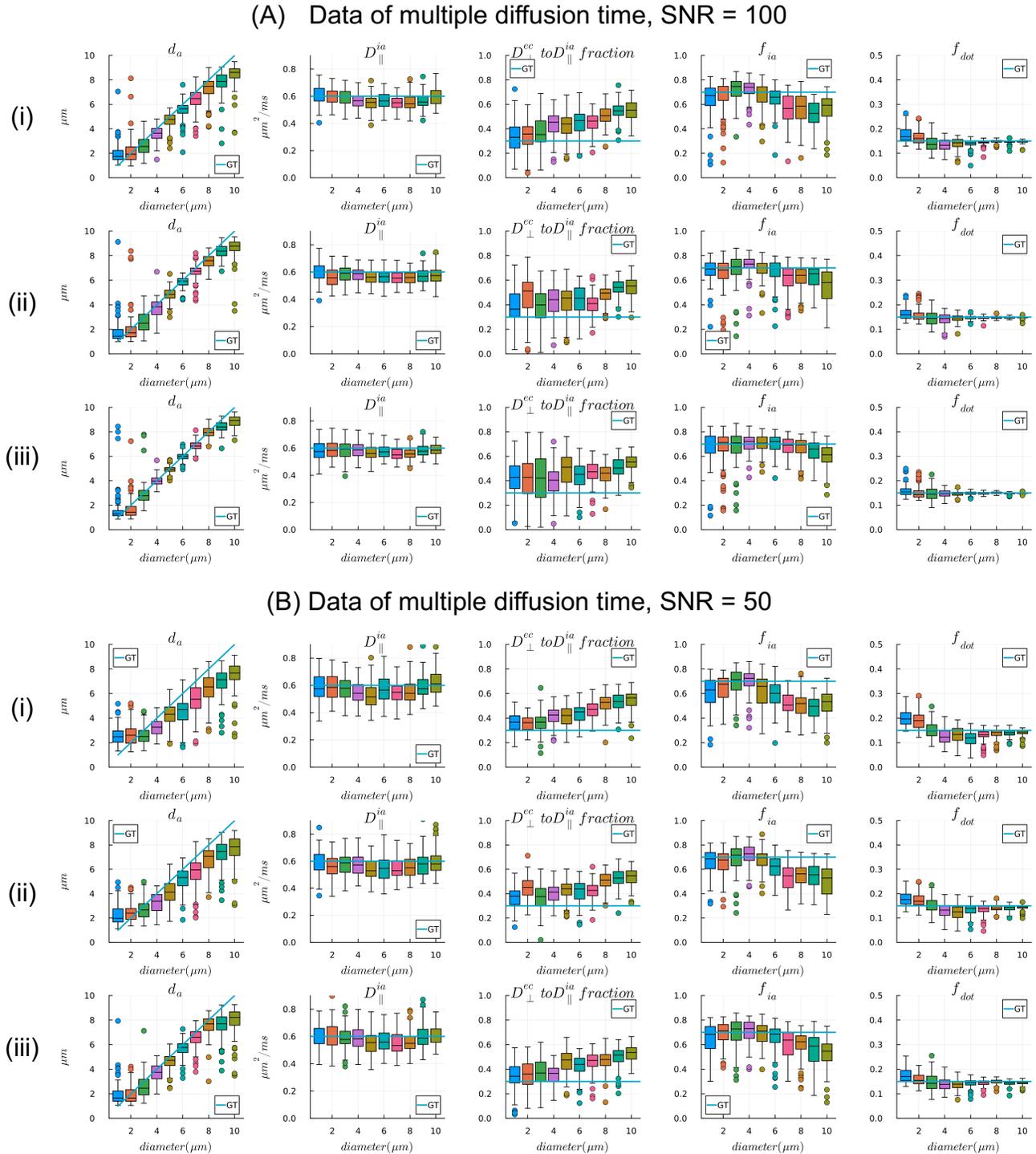

**Figure 7. Estimates of axon diameter from multi-diffusion time data** when (A) SNR = 100 and (B) SNR = 50 for the spherical mean signal. (i) Data combining two shorter diffusion times: $\delta/\Delta$ = 9.6/12 ms and $\delta/\Delta$ = 11/15.2 ms; (ii) Data combing two longer diffusion times $\delta/\Delta$ = 11/15.2 ms and $\delta/\Delta$ = 11/21 ms; (iii) Data combining all three diffusion times. Parameter estimates from 100 noise realizations are shown as boxplots and ground-truth parameter values are shown as line plots.

### 3.3 Neural network estimators

For a neural network estimator, the prior distribution of tissue parameters used for training the model will affect the posterior distributions of parameter estimates. Ideally an uninformative prior such as those used in the MCMC sampler will ensure that the estimates are less biased by prior knowledge and thus truly reflect biological properties. However, due to practical difficulties in parameter estimation, e.g., with limited data and low SNR, previous studies have used strategies such as fixing certain parameters and using narrower priors to increase the robustness of estimates. The Microstructure.jl package is designed to investigate these strategies flexibly, as the parameters to estimate in a model and the prior ranges and prior distributions are input arguments for training MLP models. It generates fitting evaluations for a fitting strategy given the protocol and SNR level in the data.

#### 3.3.1 Two-compartment SMT model

**Datasets.** We used the minimally preprocessed datasets from the WashU/UMinn young adult Human Connectome Project (HCP) (Glasser et al., 2013; Van Essen et al., 2012) to demonstrate the fitting of a Stick-Zeppelin model appropriate for WM with the SMT. The HCP data were acquired at a 1.25 mm isotropic resolution with diffusion weightings of $b$=1000, 2000, and 3000 s/mm$^2$, 90 directions per shell and 18 $b$=0 s/mm$^2$ images interspersed uniformly between the DWIs, resulting in a total of 288 measurements. We used datasets from 6 subjects for evaluating the variation of estimates.

**Experiments and Results.** The SMT model with single TE data is characterized by the intra-axonal signal fraction $f_{ia}$, intra-axonal parallel diffusivity $D_\parallel^{ia}$, extra-cellular parallel diffusivity $D_\parallel^{ec}$ and extra-cellular perpendicular diffusivity $D_\perp^{ec}$. This two-compartment SMT model (Kaden et al., 2016a) originally assumed $D_\parallel^{ec} = D_\parallel^{ia} = D_\parallel$ and linked $D_\perp^{ec}$ through a model of tortuosity as $D_\perp^{ec}$ = (1- $f_{ia}$) $D_\parallel^{ec}$, thus reduced the model parameters to $f_{ia}$ and $D_\parallel$. Here, we removed the tortuosity assumption and estimated $f_{ia}$, $D_\parallel$ and $D_\perp^{ec}$ from all three b-shells of the HCP data; $D_\perp^{ec}$ was represented as $D_\parallel^{ec} = D_\parallel$ multiplied by a fraction between [0, 1]. **Figure 2(iv-v)** shows example code for setting up this model in Microstructure.jl.

We generated 30,000 training samples with tissue parameters sampled from prior ranges of [0, 1] for $f_{ia}$ and the $D_\perp^{ec}$ / $D_\parallel$ fraction, and [1, 3] $\mu$m$^2$/ms for $D_\parallel$. We used uniform distributions for sampling the $f_{ia}$ and the $D_\perp^{ec}$ / $D_\parallel$ fraction. We tested both a uniform and a more informative Gaussian distribution for the $D_\parallel$, with a mean value of 2 $\mu$m$^2$/ms that is appropriate for WM tissue. We generated signals for the training samples using the forward model and the HCP imaging protocol and added Gaussian noise to the signals to generate noisy measurements.

We first evaluated the estimations on the synthetic training data with both the uninformative and informative prior distributions for the $D_\parallel$ (**Figure 8**). The Gaussian prior for $D_\parallel$, with a mean value representative of WM tissue, improved the precision and reduced the uncertainty for the diffusivity measures. The estimation accuracy of $f_{ia}$ was not affected much by the prior distributions of diffusivities. **Figure 9** shows the parameter maps that we obtain when applying the trained models from **Figure 8** to the HCP data, where the high uncertainty of $f_{ia}$ caused by CSF contamination is highlighted by the high standard deviations in the posteriors. The reduced uncertainty of $D_\parallel$ resulting from using a Gaussian prior is also evident from the uncertainty map. Distributions of parameter estimates in the WM from 6 subjects show high consistency.

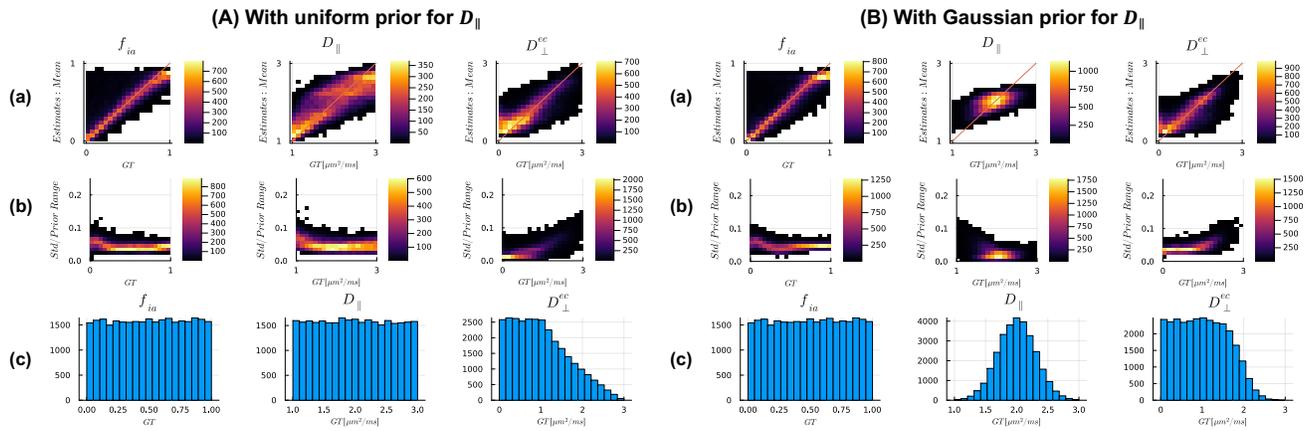

**Figure 8. Effect of prior distribution on SMT model fitting with neural network estimator**. Data were synthesized assuming the acquisition parameters of the HCP protocol and either uniform (A) or Gaussian (B) prior distribution for $D_\parallel$. Samples of $D_\perp^{ec}$ were generated by multiplying $D_\parallel$ by random, uniformly distributed numbers between [0, 1]. (a) 2D histograms of ground truth (GT) values vs. estimates (posterior mean); (b) 2D histogram of GT vs. the ratio of the standard deviation (Std) of the posterior over the prior range; (c) distributions of parameters in the training samples/ground truth parameters.

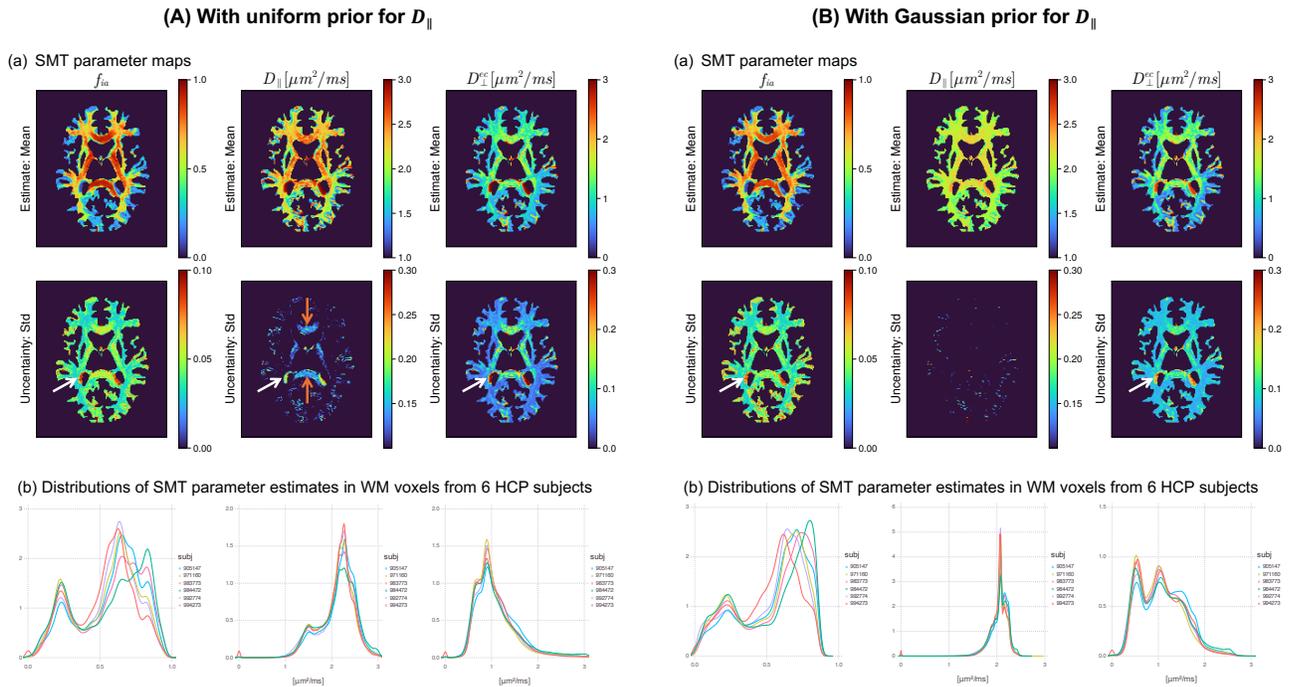

**Figure 9. SMT model fitting on in vivo human data.** The trained models with (A) uniform and (B) Gaussian priors were used for inference on 6 randomly selected datasets from the WashU/UMinn young adult HCP. (a) Mean and standard deviation maps for estimates of different microstructural parameters from one subject. The high uncertainty caused by CSF contamination in $f_{ia}$ is highlighted by the white arrows in both (A) and (B). The reduced uncertainty of $D_\parallel$ when using a Gaussian prior is highlighted by the orange arrows. (b) Probability density of the estimates in WM shows high overlap across 6 subjects.

### 3.3.2 Three-compartment SANDI model

**Datasets.** We used high b-value data acquired on the MGH Connectome 1.0 scanner (Tian et al., 2022) to demonstrate fitting the Stick-Sphere-Iso model of SANDI (Palombo et al., 2020). The datasets were acquired at two diffusion times, Δ = 19 or 49 ms, diffusion-encoding gradient duration δ = 8 ms, 8 different b-values for each diffusion time (50, 350, 800, 1500, 2400, 3450, 4750, and 6000 s/mm² for Δ = 19 ms; 200, 950, 2300, 4250, 6750, 9850, 13,500, 17,800 s/mm² for Δ = 49 ms), 32 (for b < 2400 s/mm²) or 64 (for b >= 2400 s/mm²) diffusion encoding directions uniformly distributed on a sphere. We used a subset of the data with shorter diffusion time (b = 0, 350, 800, 1500, 2400, 3450, 4750, and 6000 s/mm² and Δ = 19 ms) for SANDI, considering that the assumption of no water exchange may be invalid at Δ = 49 in the GM (Palombo et al., 2020). We used datasets from 3 subjects with a scan and rescan to evaluate the stability of estimates.

**Experiments and Results.** The free parameters in the SANDI model are the intra-soma signal fraction $f_{is}$, soma radius $R_s$, intra-neurite signal fraction $f_{in}$, intra-neurite parallel diffusivity $D_\parallel^{in}$ and diffusivity of the isotropic extra-cellular space $D_{ec}$. **Figure 2(vi)** shows example code for setting up this model in Microstructure.jl.

We generated 60,000 training samples with tissue parameters uniformly sampled from prior ranges of [2, 12] $\mu$m for $R_s$, [1.5, 2.5] $\mu$m²/ms for $D_\parallel^{in}$, and [0.5, 3.0] $\mu$m²/ms for $D_{ec}$. The compartment fractions were sampled from a Dirichlet distribution with equal concentration parameters for the three compartments. We generated signals for the training samples using the forward model and added Gaussian noise to the signals to generate noisy measurements. The noise level was set based on the temporal SNR of the b=0 images and the number of measurements used for direction averaging.

First, we evaluated the quality of SANDI model fitting on the synthetic training data as described above (**Figure 10**). While other model parameters can be accurately estimated, the $D_\parallel^{in}$ cannot be estimated with the evaluated protocol, and the estimates tend to be biased towards the mean values of the prior distribution. **Figure 11** further shows the parameter estimates when applying the trained model to the human data. The uncertainty map of $f_{in}$ features WM regions in the genu of corpus callosum with very low $f_{in}$ estimates, which could potentially be outliers. The uncertainty maps of all parameters also feature a GM region with atypical values in the $f_{in}$, $R_s$ and $D_{ec}$. The distributions of parameters in GM regions including caudate and putamen show high consistency among the 6 scans.

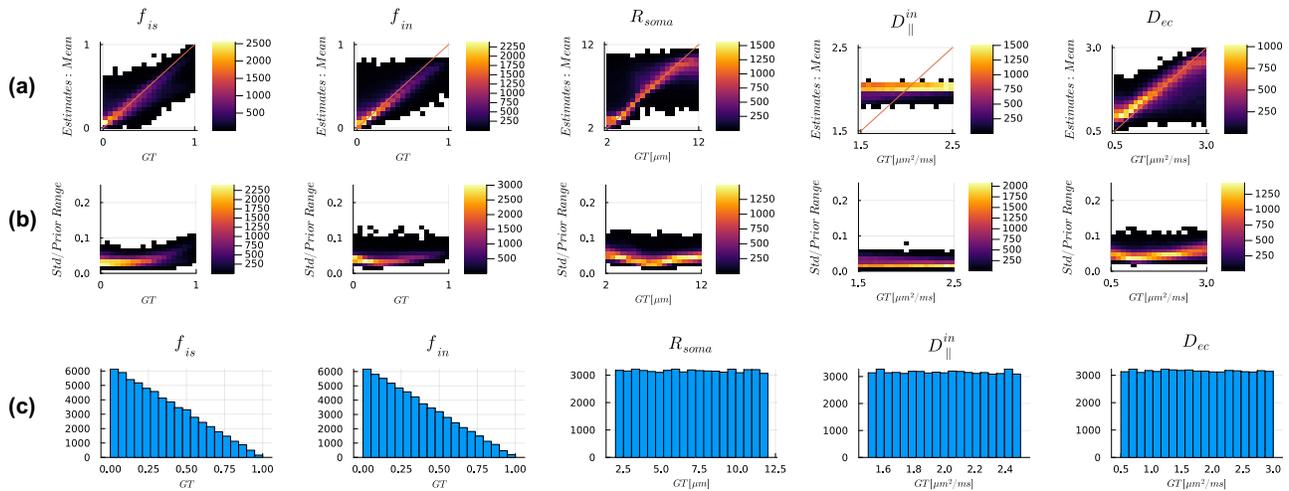

**Figure 10. Fitting evaluations of the SANDI model on synthetic training data with uninformative priors.** (a) 2D histograms of ground-truth labels vs. estimates; (b) the 2D histogram of ground-truth labels vs. the standard deviation of posteriors relative to used prior range; (c) the distributions of training labels. The signal fractions of the three tissue compartments are summed to 1 and follow the Dirichlet distribution. This evaluation suggests that the protocol is not sensitive to $D_\parallel^{in}$.

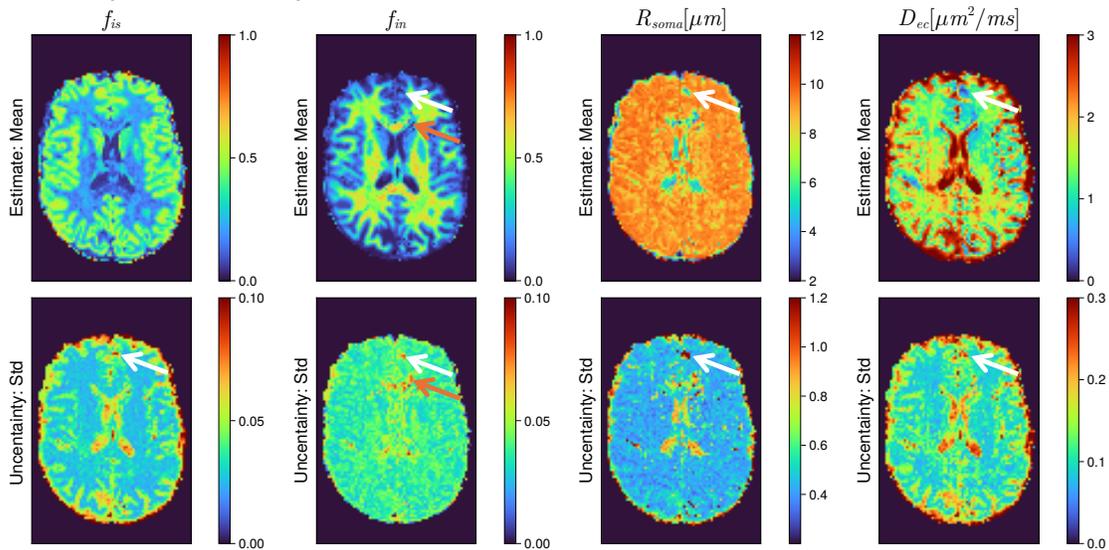

(a) SANDI parameter maps

(b) Distributions of SANDI parameter estimates in the putamen and caudate from 6 scans

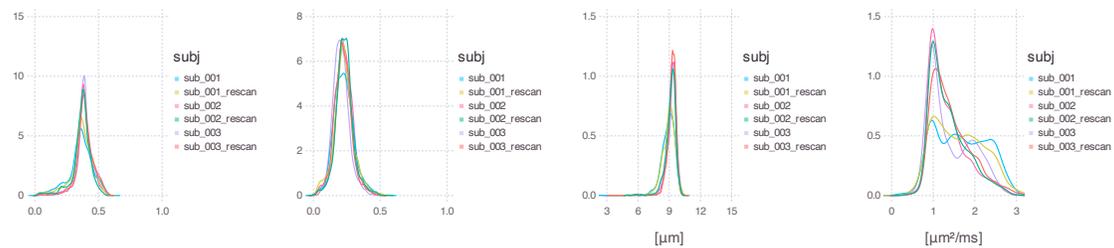

**Figure 11. SANDI model fitting on in vivo human data acquired with ultra-high gradient strength.** Only the four parameters that can be estimated robustly, based on the evaluation of **Figure 10**, are shown here. **(a)** Parameter maps from one subject. The WM region highlighted by orange arrows in the genu of corpus callosum with low $f_{in}$ values shows high uncertainty, suggesting potential outlier estimates. The GM region highlighted by white arrows showing atypical values in the $f_{in}$, $R_s$ and $D_{ec}$ exhibits high uncertainties in all the parameters. **(b)** Probability density of estimates in GM ROIs (caudate and putamen) from 6 scans.

### 3.4 Processing times

Table 1 summarizes the computation time for different tasks introduced in sections 3.2-3.3. Most analyses in this study were performed on a laptop equipped with an Apple M2 chip. On this system, the two-stage MCMC sampling required approximately 0.24 seconds per voxel using 8 threads. Training the neural network models for SMT and SANDI on the CPU took less than 10 seconds, indicating that GPU acceleration is not necessary for training small networks of this scale such as those demonstrated for SMT and SANDI.

Considering that some compartment models, such as MTE models, will have higher dimensional inputs and outputs, leading to larger network models to train, we also tested training time for a network model with more units per hidden layer (150 vs. 48) and a larger training dataset (500,000 vs. 60,000 samples) than the one demonstrated for SANDI. Training this larger network model took approximately 6.5 minutes on the laptop and 18 minutes on a Linux workstation using the CPU. On the workstation, GPU acceleration significantly reduced training time for the larger model.

Table 1. Processing times.

|  | 2-stage MCMC (task 3.2; single-diffusion time data) | Training for SMT (task 3.3.1) | Training for SANDI (task 3.3.2) | Training larger model |
|---|---|---|---|---|
| Macbook Air laptop with Apple M2 chip (16 GB) | 0.95 s/voxel (CPU; single thread) 0.24 s/voxel (CPU; 8 threads) | 2.8 s/model (CPU) | 8.1 s/model (CPU) | 6.5 min/model (CPU) |
| Workstation with Intel Xeon w7-3455 processor and NVIDIA RTX 6000 Ada Generation graphics card (48 GB) | 1.17 s/voxel (CPU; single thread) 0.15 s/voxel (CPU; 20 threads) | 4.3 s/model (CPU) | 13.4 s/model (CPU) | 18.5 min/model (CPU)<br><br>6 min/model (GPU) |

## 4. Discussion

We introduce Microstructure.jl, a package for quantifying tissue microstructure and composition across several compartment models, applicable to data acquired with typical research scanners or with high-performance/pre-clinical scanners, in vivo or ex vivo. We demonstrate some key use cases in optimizing protocols, evaluating model fitting accuracy and precision, and computationally efficient processing of high-resolution data. Microstructure.jl is included in the general Julia package registry and can thus be easily installed and updated through Julia's built-in package manager. All versions and releases can be accessed, and we maintain a website documenting its features, API manuals, and tutorials.

**4.1 Compartment models and tissue parameters**

The current release includes five tissue compartment models: Stick and Cylinder for representing axons/dendrites in typical and high-gradient data, Zeppelin for representing anisotropic extra-cellular environments, Sphere for representing cell bodies, and Iso for representing isotropic diffusion environments such as free water, extra-cellular or dot compartments. Depending on the tissue type (GM/WM), a biophysical model combines two or more compartments and estimates as many of the parameters of interest as possible with the available data.

Our demonstrations assessed several models, i.e., ExCaliber, SMT, and SANDI, which require data acquired with different protocols to achieve robust estimation of the relevant microstructure parameters. As suggested by previous studies (Andersson et al., 2022; Drobnjak et al., 2016; Fan et al., 2020; Nilsson et al., 2017; Veraart et al., 2020), the ExCaliber model requires data of ultra-high gradient strength to gain sensitivity to smaller axons most abundant in the WM. Such acquisitions are feasible only on high-performance scanners with ultra-high gradient strengths. In a separate study, we introduced this model in detail (Gong et al., 2025) and demonstrated protocol design and fitting evaluation using the toolbox presented here. This evaluation suggests that shorter diffusion time and higher gradient improves the sensitivity range of axon diameter index and that including data of multiple short diffusion times reduces estimation bias for small axons and improves estimation precision dramatically.

The SMT is a standard model with Stick and Zeppelin compartments, where the effect of fiber orientation on the dMRI signal is factored out by extracting the PA signal. It is appropriate for modelling the WM and can be fit to data acquired on widely available scanners. In our demonstration, we constrained $D_{\parallel}^{ec}$ to be equal to $D_{\parallel}^{ia}$ so that the free parameters could be fit using data with only 3 b-shells. Unconstrained estimation of the standard model parameters generally needs data acquired with multi-dimensional diffusion encoding to resolve parameter degeneracy (Coelho et al., 2022). Such data are not available in typical large-scale studies such as the HCP. However, the available multi-b-shell data could be used to extract higher order rotational invariants other than just spherical mean. This could provide additional data points in areas exhibiting anisotropy, e.g., voxels with coherent fibers, thus improving parameter fitting (Novikov et al., 2018). Our package is designed to be extendable, e.g., to include other diffusion encodings and higher-order terms from input data in the future.

The SANDI model was originally demonstrated using data from the MGH Connectome 1.0, i.e., a high-performance scanner with ultra-high gradient strength. However, recent evaluation of SANDI reproducibility suggests that it is practical on data from performant clinical scanners such as the Siemens Prisma (Schiavi et al., 2023). In this study, we demonstrated SANDI fitting on in vivo human data acquired with high b-values up to 6 ms/$\mu$m² and short diffusion time (<20 ms), as in the original study. For users interested in applying the SANDI model with Microstructure.jl, a fitting evaluation like the one in **Figure 9**, matching the acquisition protocol and SNR level to their own data, is recommended.

## 4.2 Estimators
### 4.2.1 MCMC

The effectiveness of MCMC sampling for approximating posterior distributions can be affected by various factors. On the one hand, chain convergence is affected by sampling parameters such as step sizes in parameter proposal distributions, length of chain etc. On the other hand, the acquisition protocol or SNR of the input data may limit sensitivity to the sampled model parameters. We have not tested or optimized sampling parameters for each model exhaustively. However, our package includes chain diagnostic tools (Vehtari et al., 2019), which allow the users to optimize MCMC sampling parameters for their data protocol and parameters of interest. Example diagnostics are given on our documentation website.

The length of chain is the major factor that affects computation time; implementing shorter chains than those currently demonstrated, as recommended by (Harms and Roebroeck, 2018) will reduce computation time for whole-brain datasets. Moreover, using a more informed initial guess as a starting point, rather than a random one, could potentially accelerate the sampling process. This initial guess can be provided for a whole-brain dataset with minimal additional computational cost by leveraging (potentially biased) neural-network estimators (Gong et al., 2023a).

### 4.2.2 Neural network

We demonstrate the use of neural network estimators for different models and highlight how one can evaluate model fitting on synthetic data and compute uncertainty measures that could be helpful for interpreting results from real data acquired with a certain protocol. While parameter priors can be optimized to improve estimation of certain parameters, we have focused on demonstrating this flexibility rather than showing optimized results.

For neural network estimators, the package offers full flexibility in terms of the size of the MLP, how the MLP is trained, and what data the MLP is trained on. In our experience, small MLPs, e.g., 3 hidden layers with 32 or 48 neurons in each layer, are sufficient for fitting the SMT and SANDI models. For more complex tasks with higher-dimensional parameter spaces, e.g. diffusion-relaxometry models, increasing the size of the MLP could improve the network's ability to fit the data. Increasing the depth of the MLP has been suggested to be more effective than increasing its width (Shwartz-Ziv et al., 2024).

Based on the size of the MLP, the number of training samples can be chosen to approximate the maximum capacity of the MLP to fit data. Based on practical evaluations (Shwartz-Ziv et al., 2024), small MLPs in easier tasks can usually fit more training samples than the number of parameters in the MLP model. However, for larger MLPs, the number of samples the model can fit is usually lower than the number of parameters; it is thus not recommended to include a large number of training samples due to increased training time. We have added functionality to recommend a number of training samples equal to the number of parameters multiplied by a factor of 50 (between 10-100) for smaller MLPs. In other circumstances, we recommend trying a number of training samples slightly higher than the number of training parameters.

As we have demonstrated, the prior distributions of the training data have a significant impact on parameter estimation. Previously, we have generated samples using prior distributions of tissue parameters estimated from typical human brains. This allows us to synthesize DWI signals with realistic structures for training convolutional neural networks (CNN) to estimate fiber orientation distribution (Lin et al., 2019). While priors including brain structure benefit CNN training, they introduce imbalanced distributions of parameters in the parameter space and result in biased estimates, particularly in regions where the true parameter values are underrepresented in the training sample. This issue was demonstrated in our diffusion kurtosis imaging study (Li et al., 2019) and highlighted in a simulation study (Gyori et al., 2021). In Microstructure.jl, the training distributions can be flexibly specified to improve estimation for certain parameters.

We used the root-mean-squared error from all model parameters as the loss function during training, where we also scaled the parameters to similar range (below 1) to maintain more balanced optimization errors across different model parameters. Other loss functions, such as mean-squared errors or user-defined functions, can also be tested and compared for different tasks.

### 4.3 Limitations and ongoing work

Parameter posterior distributions contain rich information that the current release of Microstructure.jl does not explore fully. A recent study fits parameter posterior distribution to Gaussian mixture models for the evaluation of parameter degeneracy, estimation uncertainty, and ambiguity at voxel level (Jallais and Palombo, 2023). Currently, we only save mean and standard deviations from posteriors due to computation time and memory constraints. In future updates, we will add functionality to save more summary and diagnostic metrics from posterior distributions when processing larger datasets.

The more advanced diffusion-relaxometry models that can be fit with Microstructure.jl require extensive validation and evaluation, which was beyond the scope of this paper. In an initial demonstration, we fit the multi-TE SMT model on dMRI data with two TEs and two b-values that can be acquired on widely available research scanners (Jun et al., 2024). In addition, we are evaluating MTE-SANDI in a separate study with multi-echo and high b-value dMRI data acquired on a high-performance scanner. We hypothesize that training data synthesized using uniform and univariant parameter distributions in a high-dimensional parameter space might introduce unrealistic or degenerate tissue parameter configurations into the training data, thus biasing parameter estimation. While the package already has the capability to perform such fitting and evaluation, further experiments are needed to optimize the data protocol and neural network estimator settings. We will further validate the toolbox and give instructions for such tasks.

## 5. Conclusion

We introduce Microstructure.jl, a Julia package featuring fast and probabilistic parameter estimation for microstructural compartment models. The package offers a consistent and extendable framework that is designed to be user-friendly both for users who would like to fit the currently supported models to their data, and for developers who would like to add support for new models.